\documentclass[preprint, 12pt]{aastex}
\usepackage{graphicx}

\begin{document}        

%% Title page

\title{The Angular Expansion and Distance of the Planetary Nebula BD+30$^\circ$3639}
\author{Jianyang Li and J.~Patrick Harrington,}
\affil{Department of Astronomy, University of Maryland, College Park, MD 20742}
\email{jyli@astro.umd.edu, jph@astro.umd.edu}
\and
\author{Kazimierz J. Borkowski}
\affil{Department of Physics, North Carolina State         
University, Raleigh, NC 27695}
\email{kborkow@unity.ncsu.edu}        

%% Abstract and subject headings

\begin{abstract} The WFPC2 camera aboard the {\it Hubble Space Telescope} was
used to obtain images of the planetary nebula \mbox{BD+30$^\circ$3639} at
two epochs separated by 5.663 years. The expansion of the  nebula in the
H$\alpha$ and [\ion{N}{2}] bands has been measured using several methods. 
Detailed expansion maps for both emission lines were constructed from nearly 
200 almost independent features. There is good agreement between the 
(independent) H$\alpha$ and [N~II] proper motions. There are clear deviations 
from uniform radial expansion, with higher expansion rates in regions where 
the shell is faintest, such as the south-west quadrant.

The Space Telescope Imaging Spectrograph (STIS) was used to obtain echelle
spectra in the C~II] $\lambda$2326 multiplet and the [O~II] $\lambda$2470
doublet, providing well-resolved expansion velocities at two position angles.
From the C~II] lines, we find that the central velocity split is $\pm$36.3
km s$^{-1}$ at a position angle of 99$^\circ$, and $\pm$33.5 km s$^{-1}$
at  p.a. 25$^\circ$. The fainter [O~II] doublet does not appear to differ
from the C~II] multiplet.

To determine the distance of BD+30$^\circ$3639 by comparison of the angular
expansion and the spectroscopically determined radial expansion, we must address
the problem of the three dimensional shape of the nebula. We measured the angular
expansion along the position of the 99$^\circ$ echelle slit, finding displacements
of 4.25 mas yr$^{-1}$ at the shell edge (2\farcs47 from the center). If the
nebula were spherical, this would imply a distance of 1.80 kpc. But there is
evidence that the nebula is elongated along the line of sight, which suggests
that the actual distance is less. Radio continuum images from 5 and 15 GHz
VLA observations provide information on the extent of the radial elongation.
We fit the radio brightness variation and the echelle data by approximating
the nebula as an ellipsoid, also making use of the ground-based echelle spectra
reported by Bryce \& Mellema (1999, MNRAS, 309, 731). Our model has an axial ratio
of 1.56, is inclined to the line of sight by 9\fdg7, and exhibits an expansion 
in the plane of the sky which is 2/3 that in the radial direction, leading to a
distance of \mbox{$1.2$ kpc}. Not all the kinematic data fits this simple model,
so the distance must still be regarded as uncertain.

Based on the recent model atmosphere of \citet{crow02}, a distance of 1.2 kpc 
implies a stellar luminosity of 4250 L$_\odot$.
The kinematic age of the nebula, $\theta/\dot{\theta}$, varies 
somewhat from region to region. A good average value is 800 years, while the 
expansion along the position of the 99$^\circ$ echelle slit gives about 600 
years.

\end{abstract}        
\keywords{planetary nebulae: individual (BD+30$^\circ$3639) --- stars: distances}      
        
%% Text

\section{INTRODUCTION}

Understanding the fundamental properties of planetary nebulae (PNe) requires 
knowledge of their distances. For planetary nebulae, because of the absence
of a standard quantity common to all, there is no well-calibrated standard 
distance scale and distances estimated from different independent methods often
disagree with each other. One way to determine the distance of a planetary 
nebula is to measure its angular expansion during a period of time and combine 
this with its Doppler expansion velocity. This method has been used
for both VLA (Very Large Array) radio maps \citep{haj93, haj95, haj96, kaw96},
and optical images from the {\it Hubble Space Telescope} ({\it HST})
\citep{red99}.

\mbox{BD+30$^\circ$3639} (hereinafter, ``BD+30'') is a rapidly evolving,
chemically inhomogeneous nebula \citep{wat98}. It is one of the few nebulae
with detected X-ray flux from its bubble of shocked stellar wind \citep{kre92,
arn96, gue00, lea00, kas00}. This justifies extensive observations  
of this object. To understand the rapid chemical evolution of BD+30, we need to
know the mass and luminosity of its central star, and this requires a
reasonably accurate distance. The earliest expansion distance of BD+30 was 
obtained by \citet{mas89} from VLA maps, who gave a result of
$2.8_{-1.2}^{+4.7}$ kpc. \citet{haj93} used different two-epoch VLA
observations, resulting in a distance of \mbox{$2.68\pm0.81$ kpc}.
\citet{kaw96} combined the observations used by \citet{mas89} with a new set of
data obtained in 1993, and derived the most accurate measurement at that time
of $1.5\pm0.4$ kpc.

To obtain a more accurate
determination of the distance, and details of its expansion, we used the
two-epoch optical observations of {\it Wide Field Planetary Camera}
({\it WFPC2}) on board the {\it HST} to measure the expansion of this nebula.
The total expansion of BD+30 was estimated to be about 15 mas between the
first and second epochs, which is about 1/3 of the {\it Planetary Camera}
({\it PC}) 0\farcs0455 pixel. This tiny displacement in diffuse objects can be
measured because of the high resolution of {\it WFPC2} images \citep{cur96},
and we can do even better for a number of diffuse knots sharing a common
proper motion. 

Much evidence shows that BD+30 is not actually a spherical object
\citep{mas89, bac91, shu98, bry99, bac00}. 
We thus use an ellipsoidal kinematic model of BD+30 to combine the angular
expansion with the radial expansion velocities from spectroscopic observations
to find the distance and kinematic age of this nebula.

\section{Observations}

{\it HST} narrow-band images from our GO programs 5403 and 8116 were used for 
this study (Table \ref{observtable}); the observations were separated by 5.663 
years. For both epochs, images through F656N (H$\alpha$) and F658N ([\ion{N}{2}])
narrow-band filters were obtained. 
In the first epoch there were two images for each band  with the same
pointing but different exposure times. The images with longer
exposure time were saturated in places. For the second epoch, there were
three images for each band, with the same exposure time and orientation, but
slightly shifted by fractional pixels with respect to each other for the
purpose of drizzling.
Since the orientation differs for the two epochs, the 
images had to be aligned to a common center and orientation.

\section{Data Reduction}

The data reduction includes cosmic ray rejection, drizzling, and image
alignment. Other adjustments or corrections like geometric distortion
correction, saturated pixel masking, were also done during the three main
steps.

Because the 2nd epoch H$\alpha$ and [\ion{N}{2}] images were dithered 
cosmic rays cannot be removed by the commonly used IRAF task
{\it crrej}. However, \citet{mut97} and \citet{fru98} introduced a method to
remove cosmic rays from a group of dithered {\it HST} images, as well as an
IRAF package {\it drizzle} for this purpose. We first
used IRAF task {\it cross\_driz}, {\it shiftfind}, and {\it imshift} to
find the relative displacement between images, and align the images. Then, we
used {\it crrej} to combine the aligned images to produce 
cosmic-ray-free images. Finally, {\it driz\_cr} was used to
create cosmic ray masks, which were used in the drizzling step.

Using the IRAF package {\it drizzle}, three 2nd epoch H$\alpha$ images and
three 2nd epoch [\ion{N}{2}] images with {\it PC}
image resolution 800$\times$800 were combined into images with resolution
1600$\times$1600, respectively. To compare images from the two epochs, the old
(the 1st epoch) images were also mapped into the same 1600$\times$1600 grid by the
{\it drizzle} package, and cosmic rays were removed by the same strategy as for
new (the 2nd epoch) images. It is easy to tell that some features are sharper in the new
drizzled images than in the old images (which were merely mapped onto finer
grids). The drizzled H$\alpha$ image is shown in Fig. \ref{drizzled_Halpha}.

To perform the comparison between new images and the old ones, images had to
be aligned in displacement and position angle. The relative rotational angle
and shift between two epochs of images were found by IRAF task {\it
cross\_driz}, {\it rotfind}, and {\it shiftfind}, iteratively. The most ideal
and simplest situation is that the expansion is isotropic, so that it looks
symmetrical. We took this as the 0th order approximation, and did the alignment
by trail-and-error strategy with the following measurements to make the
measured average expansion along radial lines as symmetrical as possible.

\section{Expansion Measurement}

By blinking the aligned images taken at two different epochs back and forth,
the expansion can be seen clearly by eye. The expansion in 1600$\times$1600
images is estimated about 0.5 to 1 pixel (or 0.25 to 0.5 PC pixels). The actual 
measurements were performed using several different independent methods.

\subsection{Methodology}

Basically, there are two ways to measure the expansion; one is magnification,
which is whole-image-based, the other is to measure the shifts of 
individual features.
If the expansion is spherically symmetric, the new image should be a magnified
version of the old image. This method can be applied to the whole image, or along sectors 
in particular directions from the image center. 

Because the old (first epoch) images have many saturated pixels, while the
new (second epoch) ones do not, we reduced the new images to fit the
old images, which avoids any manipulation of the saturated-pixel masks. 
The reduction factor was determined by minimization of the square of the
difference image rather than by cross-correlation, which we found less 
sensitive. We estimate the uncertainty at 25\% for the global
magnification factor and for the magnification factors along radial lines.

However, since the expansion of the nebula may be asymmetrical, the magnification
method cannot provide the details of the angular expansion. Thus we made
measurements of individual features and took the results of the magnification
method as a consistency check.
Compared to other PNe such as NGC~6543, BD+30 is more diffuse, so it is not
possible to measure small regions of only a few pixels in size. To determine
the best size of regions for shift measurements, 
we tested the effects of square size on some randomly selected features
by plotting the curve of measured shifts against the size of the region.
We found that when the square size was around 20 pixels (i. e., 10 PC pixels), 
shifts were the most insensitive to the size of the region. We therefore
adopted 20 $\times$ 20 pixel squares for the measurements.

The basic compare-and-fit strategy used here is just like what is used to
produce a cross-correlated image. First, move one image by, for example,
\mbox{$\Delta x$} pixels in the $x$ direction and \mbox{$\Delta y$} pixels in
the $y$ direction. Then compare and quantify the difference by 
summing up the squares of differences of corresponding pixels as in a
least-squares fit. This gives the value of the point
\mbox{($\Delta x$, $\Delta y$)} in the least-squares image. Finally, find
the minimum of the least-squares image by Gaussian 2-D fitting. This position
\mbox{($\Delta x$, $\Delta y$)} gives the relative shift between two
features.

Care must be exercised in the treatment of the edges of the squares, since
almost all regions cut from the original images have non-zero edges (and the
edges of saturated pixel masks are also sharp discontinuities). Tapering is 
not satisfactory, because the regions are very small, as are the shifts;
tapering the edges will drag the measured shifts to zero. The method we 
employed was to treat the 20 pixel square, combined with the saturated pixel
mask within the square region, as a window. We first moved the second epoch
image, which contained no saturated pixels, then looked at both images through
that window, and compared the visible
parts by least squares. To simplify the process, IDL routines were developed
which can select small regions from images interactively, and calculate the
shifts automatically.

\subsection{Results}

We used the magnification method at two levels, one for whole image, another
along every radial line at 10$^\circ$ intervals from 0$^\circ$ to 360$^\circ$.
Numerical fitting for the whole image gave magnification
factors of 0.99384 for H$\alpha$ and 0.99361 for [\ion{N}{2}].

The results from radial line magnification are shown in Fig. \ref{magniflines}.
As mentioned above, one criterion of alignment was to make the angular
distribution of magnification as symmetrical as possible. Even so, we can
still notice that the measured expansion shows some structure. The
expansion is greatest near the openings in the northeastern and southwestern
quadrants. While the distribution of walls and openings is irregular,
the smallest expansion seems associated with regions where the shell
is brightest. The direction of
greatest expansion is not along the longest axis of the nebula's
image; this may be because the nebula is a tri-axial ellipsoid, with none of
the axes aligned with the line of sight, so the expansion might not be aligned
along the longest axis of the image. This might also explain why some features
near the center were measured to have inward displacements.

The shifts of nearly 200 features in the H$\alpha$ images and the [\ion{N}{2}]
images are shown in Fig. \ref{featureH} and Fig. \ref{featureN},
respectively, which illustrates the details of expansion. Again, except for 
a few features, the results from the two independent wave-bands show a high 
degree of consistency (Fig. \ref{compare}). The agreement between H$\alpha$
and [\ion{N}{2}] gives us confidence in the results.
The discrepancies between the two wave-bands may partly be due to
slight misalignment, so that the same coordinates in different images might refer
to slightly different regions. In addition, since the ratio of [\ion{N}{2}] to
H$\alpha$ emission varies from point to point within the nebula, there may
be real differences in the shifts of features seen in the two wave-bands.

The expansions of the major axis and minor axis of the optical shell are
summarized in Table \ref{exptable}. The kinematic age of the nebula, defined
as $\theta/\dot{\theta}$, varies somewhat with the method of measuring the
expansion as well as from region to region. A good average value is 800 years.
The magnification factor derived
for the whole image gives kinematic ages of 920 and 880 years for H$\alpha$
and [N~II], respectively. The magnification factors along sectors give somewhat
smaller ages, and the ages along the minor axis tend to be less than along
the major axis. Finally, expansions from feature shifts give the smallest
kinematic ages, 740-800 years, and are again smallest along the minor axis.
As discussed below, we are particularly
interested in the expansion along the position of a spectrographic slit at
position angle 99$^\circ$; the kinematic age along this direction is about
600 years.

\section{Modeling of BD+30 and Distance Measurement}

If we want to derive the distance of the nebula,
some kind of kinematic model -- implicit or explicit -- must be employed to 
relate the tangential expansion rate to the expansion velocity along the line
of sight.

\citet{bry99} recently discussed the kinematics of BD+30 based on 
ground-based optical emission-line spectra. They
measured the spectra along two slits roughly aligned to the east-west and
north-south directions through the center, which,  approximately, 
are the major and minor axes of the nebula. For the modeling of BD+30,
they used a tilted ellipsoidal shell, with a high velocity expanding
H$_2$ ring in the equatorial plane. Furthermore, they found that 
there is a difference between the low ionization and high
ionization regions of this nebula. Compared with low ionization regions, the
high ionization regions are smaller, but have much higher velocities. They 
found the expansion velocity along the line of sight to the central star to be
$28\pm1$ km s$^{-1}$ for [\ion{N}{2}] profiles and $35.5\pm1$ km s$^{-1}$ for
[\ion{O}{3}] profiles.

\subsection{The HST STIS Echelle Spectra}

Because the nebula is compact, ground-based long-slit spectra do not have the
optimum degree of spatial resolution. As part of our HST program, we obtained
STIS spectra with the E230H echelle grating using a $6\arcsec$ x $0\farcs2$ 
slit. The grating was set to include the strong \ion{C}{2}] multiplet 
$\lambda$2324.21, 2325.40, 2326.11, 2327.64 and 2328.84, as well as the 
[\ion{O}{2}] doublet $\lambda$2470.97, 2471.09. The echelle mode of STIS is
available only with the ultraviolet MAMA detector, so substantial  dust extinction
was unavoidable. Two position angles were observed, 25$^\circ$ and 99$^\circ$.
The 25$^\circ$ orientation suffers badly from extinction, especially to the north,
so we we rely primarily on the 99$^\circ$ p.a. slit. The slit was positioned so
that it passed $0\farcs4$ north of the central star to avoid contamination by
the stellar continuum. The location of the 99$^\circ$ slit on the H$\alpha$
image is shown in Fig. \ref{slitfit}. To increase the S/N ratio, the spectrum 
was shifted by 1.53{\AA} and added to itself to superimpose C~II] 2327.64  and
2326.11, the two strongest components. The results are shown in Fig. \ref{spectrum}.
Panel A is the 25$^\circ$ orientation and B1 and B2 the 99$^\circ$ slit.
In this figure, the top of the slit is to the east. The [O~II] line appears 
identical to C~II], but has lower S/N, so we do not discuss it further.

We see that the line has well-defined red and blue components, which yield an 
accurate expansion velocity. If the nebula were a uniformly expanding ellipsoid,
the echelle profile would be an ellipse -- tilted, if the ellipsoid major axis
were inclined to the line of sight. While our profile is roughly elliptical, 
there are clearly local irregularities in the expansion. Panel (B1) shows
an ellipse which fits the data reasonably well. The vertical extent of this ellipse is 4\farcs94
and the horizontal splitting (at the center) is $\pm$36.25 km s$^{-1}$. The ellipse
is tilted to the right by 5$^\circ$ from vertical. While the splitting seems to
be well determined, the tilt of the ellipse is not secure.

But basically, it is a
tilted ellipse, which indicates that a tilted ellipsoidal shell is an
appropriate model of the nebula. The lower edge is fainter due to local dust
extinction \citep{har97}. Since C~II, like N~II, is a lower stage of ionization, 
we were surprised that the expansion velocity of this line was close to
the result \citet{bry99} found, not for [N~II], but for the high ionization 
[\ion{O}{3}] line. Perhaps our higher spatial resolution gives a more accurate 
velocity, while the lower ground-based resolution fills in the profile
center, lowering the average velocity. Another consideration is that,
although the C$^+$ ions will avoid the highly ionized central regions,
the excitation level of this UV line is high, so the emission will be 
biased toward the hottest parts of the nebula.

Because the expansion velocity was only measured along the STIS slit, we 
made measurements of the astrometric shifts of features along that cut. 
According to our kinematic model, the tangential expansion is along radial 
directions, and is linearly proportional to the distance from the center.
Under these conditions, 
projections of expansion velocities on any axis are proportional to the
corresponding projections of distances on that axis. So for point
\mbox{($x$, $y$)} of the image where the center is at \mbox{(0, 0)}, its
expansion velocity \mbox{($v_x$, $v_y$)} satisfy \mbox{$v_x \propto x$} and
\mbox{$v_y \propto y$}, and it does not matter which direction you specify as
$x$ direction. Therefore, we extracted the components of the shifts along the
slit for each feature along the cut, and plotting them against the distances to
the center of the slit, fitted them to a straight line, as shown in Fig. \ref{slitfit}.
The least-squares line goes through the center, which is a good demonstration of symmetry.
Obviously, the error of the central point will be large, since its shift should
be perpendicular to the slit. A fit excluding the center three points,
which are likely to have large errors, gave almost the same line. In this way,
the expansion rate at the edge of the nebula, which corresponds to the upper
and lower edge of the spectral ellipse, was measured to be
\mbox{$4.25\pm0.32$ mas yr$^{-1}$} at $2\farcs47$ away from the center.

If the nebula were a uniformly expanding spherical shell, then the above 
tangential expansion, combined with the 36.25 km s$^{-1}$ velocity along the same
cut, implies a distance of 1.80 kpc. There is evidence, however, that the nebula
is not spherical. In the next section, we will attempt to develop a better
kinematic model.

\subsection{Shape Derived from the Surface Brightness Variation}

BD+30 is optically thick to ionizing radiation -- it is surrounded by a neutral
halo \citep{har97} -- but the photoionized shell is also geometrically thin 
(perhaps because of the pressure of the $3 \times 10^6$ K X-ray emitting gas
which fills the interior of the shell). Under these conditions, the variation in
the surface brightness across the nebula contains information about the three
dimensional structure of the nebular shell. Since it would be difficult to correct 
for the internal extinction due to dust which affects the optical images, we have
used the 5 and 15 GHz VLA radio maps described in \citet{har97} for this analysis.
A similar approach was previously used by \citet{mas89}, who modeled the 
apparent EW elongation and surface brightness variations of BD+30 in terms of
an inclined prolate ellipsoid tilted from our line of sight along the EW
direction, consistent with a limited kinematic information available at that time.

The relevant equations are developed in the Appendix. The constancy of flux in 
pie-shaped sectors radiating from the central star supports the idea of complete
absorption of the ionizing radiation. The ratio of the central surface brightness
to the total flux indicates that the nebula is elongated along the line-of-sight:
the central intensity is substantially less than would be the case for a sphere.
Unfortunately, the angular expansion vanishes near the center of the nebula; what
we need is information on the shape of nebula where we have measured the angular
expansion. We can, however, find the shape of the shell along a chosen wedge from 
the surface brightness distribution (eq. A3). We integrated the surface brightness
in 25$^\circ$ sectors chosen to overlap the region cut by our 99$^\circ$ STIS slit. 
The results are shown in Fig. \ref{shell_sect}. The values near the central star are 
very noisy due to the low signal at the wedge apex. Also, near the edge of the nebula,
the finite shell thickness invalidates the assumptions of the method. But overall, 
the shell seems to be well fit by an ellipse with an axial ratio of 1.5 to 1 along
our line-of-sight, as indicated by the dotted line in Fig. \ref{shell_sect}. This 
fit also closely matches the shell distance of 3\farcs7 in the direction of the 
central star -- the triangle in Fig. \ref{shell_sect} -- found from the central 
intensity (eq. A1).

In obtaining the shape from eq. A3 we made the assumption that the front and back
surfaces of the shell contribute equally to the surface brightness. But if we look
at the echelle profile, we see that there is a great asymmetry between the front
(blue-shifted) and back (red-shifted) branches. We might attribute this to dust extinction,
and indeed we feel that the relative weakness of the lower (west) part of the profile
is likely due to dust. But this cannot explain the relative brightness of the upper,
red-shifted side of the line, because this radiation comes from the {\it back} side 
of the nebula: this radiation must suffer at least as much extinction as the weaker 
blue side.

A slit placed across a uniformly expanding shell in the form of a tilted ellipsoid 
will result in a spectral line in the form of a tilted ellipse. (The tilt angle of 
the major axis in a plot of the line profile will vary depending upon the relative 
scales of chosen for the x-axis and y-axis. The ratio of the x displacement of the 
top of the ellipse to the width of the ellipse on the x-axis is, however, an 
invariant.)  Because the ellipse that best fits our echelle profile seems tilted, 
we considered the appearance of a tilted ellipsoidal shell. Equation A7 gives the
front and back surface brightnesses, which we found, for the parameters relevant here,
to differ by up to a factor of two. We can sum the front and back contributions and 
apply eq. A3 to the result, to simulate our analysis of the VLA data assuming symmetry.
We find that an ellipsoid of axial ratio 1.56, tilted by 10$^\circ$,
closely resembles an untilted ellipsoid of axial ratio 1.5. (Such a solution is 
shown on Fig. \ref{shell_sect} as a dashed line.) 

Since the top of our echelle spectrum seems tilted redward, this would imply an 
ellipsoid with the near side tilted down (westward). Unfortunately, with such a 
tilt, it is the front (approaching, blue-shifted) side which is brighter at the top
(east) end of the slit, just the opposite of what is observed. In an attempt to
clarify this situation, we examined the spectra presented by \citet{bry99}, who
reproduce two spectra taken with an east-west slit (close to the orientation of 
our 99$^\circ$ slit), one in the [O~III] $\lambda$5007 line and the other in the
relatively weak [N~II] $\lambda$5754 line. The [O~III] profile is clearly tilted
with the east side blueward, {\it opposite the apparent tilt of our C~II] line.} 
The [N~II] profile also seems to be tilted with the east side blueward, but
determination of the tilt is difficult in this case because the parts of the
profile corresponding to high velocities are very faint compared to the bright 
parts at lower velocities. We find that if we apply the same analysis as above
to the STIS profile, this time fitting the VLA surface brightness and the
tilt of the \citet{bry99} [O~III] profile, we get a good fit with an ellipsoid 
of axial ratio 1.56, with the near end tilted up (eastward) by 9.7$^\circ$. 
This is the same direction of tilt shown in Fig.~4 of \citet{bry99}. We
were guided by the C~II] profile in setting the central velocity and spatial
extent of this fit, since the relatively low resolution of the [O~III] profile 
provided less constraint. Fig. \ref{bryce_fit} shows this predicted profile 
superimposed on the \citet{bry99} data. 

While no simple ellipsoidal model seems able to fit these complex data, we feel
that this is the best compromise, since it (a) fits the VLA surface brightness,
(b) produces a front/back brightness ratio in the same sense as that seen in
both the C~II] and [O~III] profiles, (c) fits the central velocity seen in the
C~II] profile, and (d) fits the direction and magnitude of the tilt seen in the 
[O~III] line. The problem is that this model predicts a tilt in the opposite sense to 
that seen in the C~II] line remains. In view of the low S/N of this profile, 
however, its tilt is uncertain. Also, the fact that the ellipsoidal models are
only slightly inclined to the line of sight means that irregularities in the
shape of the shell can easily change the tilt of the line profiles.

The ultimate goal of our kinematic model is to relate the tangential expansion,
as determined by the linear fit in Fig. \ref{slitfit}, to the Doppler splitting
seen in the echelle spectrum. The model discussed above -- a tilted ellipsoid 
with a 1.56 axial ratio -- has a tangential to radial velocity ratio of 0.667.
This reduces our estimate of the distance to BD+30 to 1.20 kpc. The error 
attached to this distance is hard to estimate. The formal error in the linear 
fit to the astrometric shifts along the slit is less than 10\%, and the error
in the expansion velocity smaller yet. This implies an error of about 10\%
in the distance, but {\it only if we have an accurate kinematic model.}

It is clear that the shape and kinematics have an important but somewhat
uncertain impact on our derived distance, almost surely reducing the value
significantly below the spherical value, but by an amount that must remain 
somewhat uncertain. Progress might be made by a program of extensive kinematic
mapping with multiple narrow slits -- preferably in the infrared to minimize 
the effects of extinction by internal dust.

We should also bear in mind that all the foregoing analysis is based on the 
assumption that the expansion of the nebula is radial, and with a velocity
proportional to the distance from the center -- such an expansion will preserve
the shape of the nebula. While this is a reasonable and conservative 
assumption, it is not hard to imagine that there may be more material around
the waist of the nebula (the plane perpendicular to the major axis) which
may impede expansion in that direction. Such a situation would reduce the
derived distance to the nebula still further.

\section{Discussion}

We have used pairs of HST WFPC images separated by 5.663 years to produce 
detailed maps of the angular expansion of BD+30. Images in both H$\alpha$ and
[N~II] show similar patterns of expansion. Near the northeastern corner, the
direction deviations of expansion flow are very regular, and the magnitudes of
expansion are also close to each other. If we look at the opening near the
northeastern corner, the movements of the material in the inner part tends 
to move toward the opening, not always along the radial direction. Because
measurements were performed on small independent regions, this cannot be a
systematic error. It appears that the material in the inner part is
trying to go out through the opening, which has relatively low density, instead
of moving along a radial direction. This probably shows that the inner material
has a higher velocity than the outer shell. The west parts of the image
do not have very regular patterns over a large scale, but we can still notice
some systematic movements. For example, the expansions along the southwestern
direction going outward are almost all along one direction with same magnitude,
and there is another convergence center at the southwestern part of the image.
The velocities are particularly large at the outer edge of the southwest 
quadrant. It is interesting that the largest proper motions are approximately
along the axis of the bipolar outflow as delineated by CO "bullets" imaged by
\citet{bac00}, rather than along the (EW) major axis. Apparently, measurements 
of the angular expansion can reveal the ``true'' bipolar axis, consistent with
ground-based kinematics.

Overall, our measurements are more secure than earlier results, since the
high resolution of these images allows us to measure the displacements of
small, discrete features -- in contrast to the radio work, which could
only measure the global expansion. A comparison of proper motions listed
in Table 2 with the radio expansion measured by \citet{kaw96}
shows a good agreement, but both measurements are significantly larger
than the radio expansion determined by \citet{haj93}. The difference
between these two radio proper motion measurements arises from a correction
used by \citet{kaw96} to account for the decreasing surface
brightness of BD+30 as it expands at an assumed constant luminosity. The
present measurements confirm the validity of this correction, which must
be applied if an optically thick nebula is observed with an instrument
whose spatial resolution is comparable to the size of the object. While
the HST spatial resolution is superior with respect to the existing radio
observations, a rather remarkable agreement with measurements of
\citet{kaw96} demonstrate that ground based radio observations
can provide reliable proper motions for compact PNe.

Although our STIS spectra have rather poor S/N, their high spatial resolution
gives us a precise measure of the expansion of the brightest part of the 
nebular shell in a well defined location. This velocity data can be combined 
with the astrometric shifts along the same cut to derive a distance. Here, the 
greatest uncertainty is the shape and kinematics of the nebula: how elongated 
is the shell and how uniform and radial is the expansion? We have used the radio 
surface brightness to address the shape, but we have had to assume uniform radial 
expansion -- even though our maps of angular expansion show clear non-radial 
motions is some sectors. Still, we feel that our resulting distance determination 
of 1.2 kpc is better justified than previous distance estimates, and offers
an improvement over the most recent value of $1.5 \pm 0.4$ kpc found by
\citet{kaw96} from radio expansion measurements combined with
the ground-based kinematic data. 

At a distance of 1.2 kpc, the luminosity of the central star of BD+30 is
equal to 4250~L$_\odot$, following the recent determination of \citet{crow02} 
based on modeling of the stellar spectrum. (This is 60\% lower than the
luminosity based on the earlier model of \citet{leuen96}.) 
Most of the stellar radiation is absorbed within the nebula,
primarily by dust in the neutral halo (the 
infrared luminosity is 2000~L$_\odot$) and in the photoionized shell
(nebular models indicate that $\sim 800$~L$_\odot$ is required to account
for the observed radio continuum). The mostly neutral nebular
mass is equal to 0.13 M$_\odot$, based on the [\,C II\,] 158 $\mu$m line
luminosity derived from the Infrared Space Observatory (ISO) measurements
of this line flux \citep{liuetal01} and the 1.2 kpc distance. In
terms of its mass and luminosity, BD+30 is not exceptional among Galactic
PNe. But its central Wolf-Rayet WC star is hydrogen poor and ISO
observations revealed the presence of crystalline silicate dust in this
carbon-rich nebula \citep{wat98}, signaling a recent change from the
oxygen- to carbon-dominated chemistry. The origin of these rapid abundance
variations
is not understood at present. One promising hypothesis involves a final
thermal pulse \citep{wat98, her01} which occurred at most a few
thousand yr ago, shortly before
BD+30 left the Asymptotic Giant Branch phase of evolution.
With the fairly well determined distance, perhaps this bright
and well-observed PN can provide a stringent test of various hypotheses
invoked to explain the origin of WC stars and the presence
of abundance inhomogeneities in PNe.

%% Acknowledgements

\acknowledgements Support for this work was provided by NASA through grants 
GO-08116.01-97A and GO-08116.02-97A from the Space Telescope Science 
Institute, which is operated
by the Association of Universities, Inc., under NASA contract NAS5-26555. We
would like to thank Dr. Myfanwy Bryce for supplying the data from \citet{bry99}
used in Fig. \ref{bryce_fit}.

%% Appendix

\appendix 
\section{Appendix}

To the extent that the temperature in the nebular gas is constant, and if we
neglect the secondary effects of diffuse radiation produced within the gas, the
emission in a hydrogen recombination line or in the free-free continuum will
depend only upon the flux of stellar ionizing radiation that is absorbed in the
volume of gas. Consider a nebular shell which is optically thick to ionizing
radiation but which is geometrically thin. Then, the emission from the surface of
the shell will simply be proportional to the solid angle of the surface element
as seen from the central star. Consider an element of the shell $dS$ which is
inclined so that the angle between the normal to the surface and our line-of-sight
is $\beta$. Let $\alpha$ be the angle between the radial vector from the star and 
the normal to $dS$. Then $dS$ will appear to have a surface brightness $\Sigma$ of
\begin{equation}
\Sigma~=~\frac{F}{4\pi}~\frac{\cos \alpha}{\cos \beta}
				  ~{\left[\frac{D}{r}\right]}^2
\end{equation}

\noindent where $F$ is the total flux from the whole nebula, $r$ is the distance of
$dS$ from the central star, and $D$ is the distance from the nebula to the observer.
Near the central star, where $\cos(\alpha) = \cos(\beta)$, the ratio of $\Sigma$ to $F$ 
provides the distance of the surface from the star (in angular units: (r/D) radians).
One complication is that we see both the front and back sides and must separate these
contributions or assume symmetry. From the central surface brightness of the radio
frequency images, it is apparent that the nebula is elongated along the line-of-sight,
since the central intensity is substantially less that would be the case for a sphere. 
We find (r/D) to be at least $3\farcs4$; extrapolation to zero radius of the flux in 
rings about the center yields $3\farcs7$.

Consider a spherical coordinate system whose origin is the central star, with the z-axis 
directed towards the observer. Call the azimuthal angle $\phi$ and the angle between the
z-axis and the vector to a point on the surface $\theta$. The flux in any pie-shaped section
of the image, bounded by angles $\phi_1$ and $\phi_2$, must be $F(\phi_2 - \phi_1)/2\pi$,
because that part of the image arises from the absorption of the fraction of the stellar
flux emitted into the volume bounded by the planes $\phi=\phi_1$ and $\phi=\phi_2$.
In fact, this provides a check on our assumption that the nebula is optically thick and
that diffuse nebular radiation is not important: equal pie-shaped segments of the radio
image should have the same flux. This check is satisfied to within ~5\%, except for a
sector at p.a. $\sim 60^\circ$, where there is deficit of 11\%. (We use the
radio images because the H$\alpha$ images are affected by dust absorption/scattering.)
Unfortunately, the angular expansion vanishes near the center of the nebula; what we
want is information on the shape of nebula where we have measured the angular expansion.

Let $p$ be the (projected) angular distance of a bit of the surface $dS$ from the
star. Then the flux in the image segment between $[p,p+dp]$ and $[\phi,\phi+ d\phi]$
is equal to the product $\Sigma(p,\phi)~p~dp~d\phi$ There are two contributions to
$\Sigma$: radiation from the front and back surfaces. We assume that we can separate 
them or that they are equal. So $\Sigma$ here refers just to the front or back surface.
The emitting piece of surface is located at the (unknown) angle $\theta$ from the star.
The flux emitted by the surface is $F~d\Omega/4 \pi$, where $d\Omega$ is the solid 
angle subtended by the surface as seen from the star. $d\Omega = \sin(\theta)~d\phi~
d\theta$, so that the flux is $[F/4\pi]\sin(\theta)~d\phi~d\theta$. By equating fluxes
we can write
\begin{equation}
\frac{F}{4\pi}~\sin(\theta)~d\phi~d\theta~=~\Sigma(p,\phi)~p~dp~d\phi
\end{equation}

Consider a thin segment of the nebula between $\phi$ and $\phi+d\phi$. Integrate 
the expression above from an angle $\theta$, corresponding to some impact parameter 
$p$, to $\theta = \pi/2$, corresponding to $p_0$. ($\theta = \pi/2$ is the plane passing 
through the star and perpendicular to the line-of-sight. If the nebula is expanding
radially, then this is the plane with zero radial velocity.)  We find that
\begin{equation}
\cos \theta~=~\frac{4\pi}{F}~{\int_p^{p_0} \Sigma(p,\phi)~p~dp}
\end{equation}

The expression on the right hand side is just the integrated flux from the azimuthal
segment (from back or front shell only), divided by $F/4\pi$. We use this relation
to find $\theta(p)$. Now $p = r \sin(\theta)$ and $z = r \cos(\theta)$, so $z = p/
\tan(\theta)$. The p-z plot is the cross-section of the nebula in the $\phi$-plane.

As a simple example, let the nebula be an ellipsoid of revolution about the z axis
with the minor axis $b$ in the plane of the sky and the major axis (z-axis) $a$ along
the line of sight. Then the radius vector from the star to a point on the surface is
\begin{equation}
r~=~\frac{a b}{\sqrt{b^2 \cos^2(\theta)~+~a^2 \sin^2(\theta)}}
\end{equation}

\noindent
The angle between the z-axis and the normal to the ellipsoid surface is  
\begin{equation}
\beta~=~\arctan [(a/b)^2 \tan(\theta)]
\end{equation}

\noindent
Then from equation (1), we can show that the surface brightens is
\begin{equation}
\Sigma~=~\frac{a^2}{r^4~\cos(\theta)}
\end{equation}

\noindent
Now, $p = r \sin(\theta)$ and $\frac{dp}{d\theta}= \frac{r^3 \cos(\theta)}{a^2}$,
so that $\Sigma~p~dp~=~\sin(\theta)~d\theta$, and equation (3) is satisfied. 

If the ellipsoid is inclined to the x-axis by an angle $i$, the surface brightness
becomes
\begin{equation}
\Sigma~=~\frac{(a b)^2}{r^4}~\frac{1}{b^2 \cos(\theta) \cos(i) -
	   a^2 \sin(\theta) \sin(i)}
\end{equation}

\noindent
In this expression, $\Sigma$ is negative for values of $\theta$ corresponding to 
the back side of the shell. Even for modest $i$ and $a/b$, the effects of $\beta$   
and $r$ may combine to produce large front-to-back shell brightness ratios.

%% References

%% Figure Captions

\clearpage
\begin{figure}
\caption{Drizzled H$\alpha$ image of BD+30. The intensity
  scale is linear. North is up and east is to the left. 
  In this figure, the field of view is 9\farcs10 $\times$ 9\farcs10.
  \label{drizzled_Halpha}}
\end{figure}

\begin{figure}
\caption{Line magnification measurements of BD+30 in H$\alpha$.
  The numbers are the magnification factor minus 1 along the
  corresponding radial directions. In this and the following figures, the
  field of view is 6\farcs37 $\times$ 6\farcs37.
  \label{magniflines}}
\end{figure}

\begin{figure}
\caption{Expansions of individual features in the BD+30 H$\alpha$
  image. All numbers are in units of mas yr$^{-1}$. 
  \label{featureH}}
\end{figure}

\begin{figure}
\caption{Expansions of individual features in the BD+30 [N~II]
  image. All numbers are in units of mas yr$^{-1}$.
  \label{featureN}}
\end{figure}

\begin{figure}
\caption{The comparison of the expansions of individual features 
  in the H$\alpha$ band and in the [N~II] band. Solid lines
  represent the H$\alpha$ shifts, and dashed lines are for [N~II]. 
  White regions in the figure are masked saturated pixels.
  \label{compare}}
\end{figure}

\begin{figure}
\caption{The C~II] $\lambda$2327 line profiles of BD+30, obtained 
 with the HST STIS. Panel (A) is with the slit at position angle 25$^\circ$ 
  east of north; north is at the top. Panels (B1) and (B2) show the 
  spectrum taken with position angle of 99$^\circ$; east is at the top.  
  In panels (A) and (B1) we have overploted the best fitting ellipses. 
  The bright dots mark the centers of the ellipses.
  Panel (B2) shows the same data as (B1) without the distraction of the
  ellipse. For both positions the slit was shifted $0\farcs4$ from the 
  central star, although a trace of continuum at the top may be a broad
  stellar line scattered by nebular dust. The apparent spur seen in (B) 
  branching from the profile at about 10 o'clock is the trace of another 
  line of the C~II] multiplet.
  \label{spectrum}}
\end{figure}

\begin{figure}
\caption{The position of slit through which the STIS echelle spectrum 
  (Fig. \ref{spectrum} B)
  was taken, and the fit to the expansion. The left panel shows the slit
  orientation and width, and the measured shifts of features along the slit. 
  The right panel shows the fit of shifts along the slit against the distances
  from the center of slit. The triangles are the shifts and the square marks
  the origin. The dashed lines at $\pm2\farcs47$ mark the edges of the fitting
  elipse shown in Fig. \ref{spectrum} B1.
  \label{slitfit}}
\end{figure}

\begin{figure}
\caption{The cross-section of the nebula (arcsec) along p.a. 99$^\circ$. 
The vertical axis is in the plane of the sky, the horizontal axis is toward 
the observer. The + symbols are from equation A~3, where for $\Sigma$ we
used the average surface brightness of the radio images over a $\phi$ 
interval of 25$^\circ$. 
The triangle is from equation A~1. The dotted line is an
ellipse of 1.5 axial ratio. The dashed line is the shape that would be
inferred from equation A~3 applied to an ellipse of 1.56 axial ratio, 
inclined by 9\fdg7. See text.
\label{shell_sect}}
\end{figure}

\begin{figure}
\caption{The white curve is our model fit plotted over the
[O~III]~$\lambda$5007 line profile of Bryce \& Mellema (1999), 
Fig. 2(d). This spectrum was taken with an E-W slit in 1\arcsec\ 
seeing. We have smoothed their original data and reversed 
the y-axis to place east at the top, as in our STIS echelle 
spectra (Fig. \ref{spectrum}).
\label{bryce_fit}}
\end{figure}

%% Tables

\clearpage
\begin{deluxetable}{ccccc}
\tablecolumns{5} 
\tablewidth{0pc}
\tabletypesize{\footnotesize}
\tablecaption{Observations Used for Expansion Measurement
  \label{observtable}}
\tablehead{
  \colhead{Filter} & \colhead{Exposure} & 
  \colhead{HST P.A.} & \colhead{Date} & \colhead{Proposal} \\
  \colhead{} & \colhead{(sec)} & 
  \colhead{(deg)} & \colhead{(yyyy-mm-dd)} & \colhead{} }
\startdata
   H$\alpha$ F656N & 200,300 & 115 40 58.8 & 1994-03-06 & 5403 \\
   H$\alpha$ F656N & 3 $\times$ 160 & 249 04 05.9 & 1999-11-04 & 8116 \\
  \ [N~II] F658N & 200,300 & 115 40 58.8 & 1994-03-06 & 5403 \\
  \ [N~II] F658N & 3 $\times$ 160 & 249 04 05.9 & 1999-11-04 & 8116 \\
\enddata
\end{deluxetable}

\clearpage
\begin{deluxetable}{lcccc}
\tablecolumns{5}
\tablewidth{0pc}
\tablecaption{Expansion Rates of BD+30 in mas yr$^{-1}$. The values
  from two magnification methods came from the magnification factors times the
  distance of the edges of the major and minor axes from the center of image.
  The values from feature measurement were calculated by averaging some values
  near the edges of major and minor axes.
  \label{exptable}}
\tablehead{
  \colhead{} & \multicolumn{2}{c}{H$\alpha$} &
    \multicolumn{2}{c}{[\ion{N}{2}]} \\
  \cline{2-3}\cline{4-5}
  \colhead{} & \colhead{Major Axis} & \colhead{Minor Axis} &
  \colhead{Major Axis}& \colhead{Minor Axis}}
\startdata
  Radius & 2\arcsec.32 & 1\arcsec.84 & 2\arcsec.32 & 1\arcsec.84 \\
  Magnification (mas yr$^{-1}$) & $2.54\pm0.51$ & $2.02\pm0.40$ &
    $2.64\pm0.53$ & $2.09\pm0.42$ \\
  Line Magnification (mas yr$^{-1}$) & $2.82\pm0.56$ & $2.30\pm0.46$ &
    $2.91\pm0.58$ & $2.40\pm0.48$ \\
  Feature Shifts (mas yr$^{-1}$) & $2.89\pm0.29$ & $2.49\pm0.25$ &
    $3.07\pm0.31$ & $2.45\pm0.25$ \\
\enddata
\end{deluxetable}

\end{document}